# DERIVING CONSENSUS RANKINGS VIA MULTICRITERIA DECISION MAKING METHODOLOGY

[1]AmyPoh. AL, [2]M. N. Saludin, [1]M. Mukaidono

[1]Faculty of Science and Technology, Meiji University

[2]Faculty of Management & Defense Study, National Defense University of Malaysia

## *ABSTRACT*

**Purpose** – This paper takes a cautionary stance to the impact of marketing mix on customer satisfaction, via a case study deriving consensus rankings for benchmarking on selected retail stores in Malaysia.

**Design/methodology/approach** – ELECTRE I model is used in deriving consensus rankings via multicriteria decision making method for benchmarking base on the marketing mix model 4Ps. Descriptive analysis is used to analyze the best practice among the four marketing tactics.

**Findings** – Outranking methods in consequence constitute a strong base on which to found the entire structure of the behavioral theory of benchmarking applied to development of marketing strategy.

**Research limitations/implications** – This study has looked only at a limited part of the puzzle of how consumer satisfaction translates into behavioral outcomes.

**Practical implications** – The study provides managers with guidance on how to generate rough outline of potential marketing activities that can be used to take advantage of capabilities and convert weaknesses and threats.

**Originality/value** – This paper interestingly portrays the effective usage of multicriteria decision making and ranking method to help marketing manager predict their marketing trend.

**Type of paper:** Research Paper

**1. INTRODUCTION**

With increasing globalization, local retailers find themselves having to compete with large foreign players by targeting niche markets. To excel and flaunt as a market leader in an ultramodern era and a globalize world, the organizations must strive to harvest from its marketing strategies, benchmarking and company quality policy.

Ranking and selecting projects is a relatively common, yet often difficult task. It is complicated because there is usually more than one dimension for measuring the impact of each criteria and more than one decision maker. This paper considers a real application of project selection for the marketing mix element, using an approach called ELECTRE.

The ELECTRE method has several unique features not found in other solution methods; these are the concepts of outranking and indifference and preference thresholds. The ELECTRE method applied to the project selection problem using SPSS (Statistical Package for the Social Sciences) application.

Our contribution is to show the potential of Marketing mix model in deriving a consensus ranking for benchmarking. According to the feedback from the respondents, we dynamically rank out the best element to be benchmark.

The decision problem faced by management has been translated into our market research problem in the form of questions that define the information that is required to make the decision and how this information obtained. The corresponding research problem is to assess whether the market would accept the consensus rankings derive from benchmarking result from the impact of marketing mix on customer satisfaction using a multi-criteria decision making outranking methodology.

## 2. LITERATURE REVIEW

The project ranking problem is, like many decision problems, challenging for at least two reasons. First, there is no single criterion in marketing mix model which adequately captures the effect or impact of each element; in other words, it is a multiple criteria problem. Second, there is no single decision maker; instead the project ranking requires a consensus from a group of decision makers. (Henig and Buchanan and Buchanan et al.)

Buchanan et al. have debated that effective decisions come from effective decision process and proposed that where potential the subjective and objective parts of the decision process should be branched. The relationship between the alternatives and the criteria is portrayed using attributes, which are the objective and measurable character of alternatives. Attributes form the bridge within the alternatives and the criteria. Often, marketing management is looking and interesting on the solution rather than the outlines criteria.

Referring to the statement of Simon (1977), analysis decisions ex post cannot accurately be done due to human memory has some known biases. Through observation, we noticed that in many cases, decision is treated as a one shot game whereas most decisions are more or less repetitious. A decision maker can learn the effect of the assignment he has distributed to the weights. Likewise, the decision maker can learn to modify concordance and discordance factors in outranking methods (Roy and Skalka, 1985; Vetschera, 1986).

In the theoretical account of decision making, we remember that, the subjective and contextual data play an important role due to the prominent look-ahead component (Pomerol, 1995). Moreover, due to the rawness of the framework, particularly during the evaluation stages (Lévine and Pomerol, 1995), explanations and contextual knowledge are among the elements facilitating the cooperation, and the need to make them explicit and shared both by the system and the user (Brezillon and Abu-Hakima, 1995) and Brézillon (1996).

# 3. RESEARCH METHODOLOGY

## 3.1 Recognizance Survey

This section takes into consideration sites in Selangor area, geographical position in the center of Peninsular Malaysia, contributed to the state's rapid development as Malaysia's transportation and industrial hub, with a population of 4,736,100 (2005 estimate). The selected data collection sites are Tesco Saujana Impian Kajang, Carrefour Alamanda Putrajaya, Giant Bukit Tinggi and Mydin Kajang.

## 3.2 Research Instrument

A non-comparative Likert scaling technique was used in this survey. The questionnaire is divided into 4 sections: customer information, marketing mix model, customer perception and motivating factor. The demography variables measured at a nominal level in Section 1 include gender, ethnic, marital status, age and how often do the respondents shop at the specific retail store.

A typical test item in a Likert scale is a statement. The respondent is asked to indicate his or her degree of agreement with the statement or any kind of subjective or objective evaluation of the statement. In Section 2, a six-point scale is used in a forced choice method where the middle option of "Neither agree nor disagree" is not available. The questions comprise four attributes such as product, price, promotions, place/distribution; six questions are allocated for each of the 4Ps.

Section 3 evaluates customer's perception using the same scale as practice in Section 2 whereas Section 4, the last part of the questionnaire measure the factor that motivates respondents the most to patronize the specific retail store using the nominal measurement. Simple random sampling technique is used in the research.

**3.3 Illustration of Research Framework**

FIGURE 1. Attribute – 4P's – Retail Stores Mapping

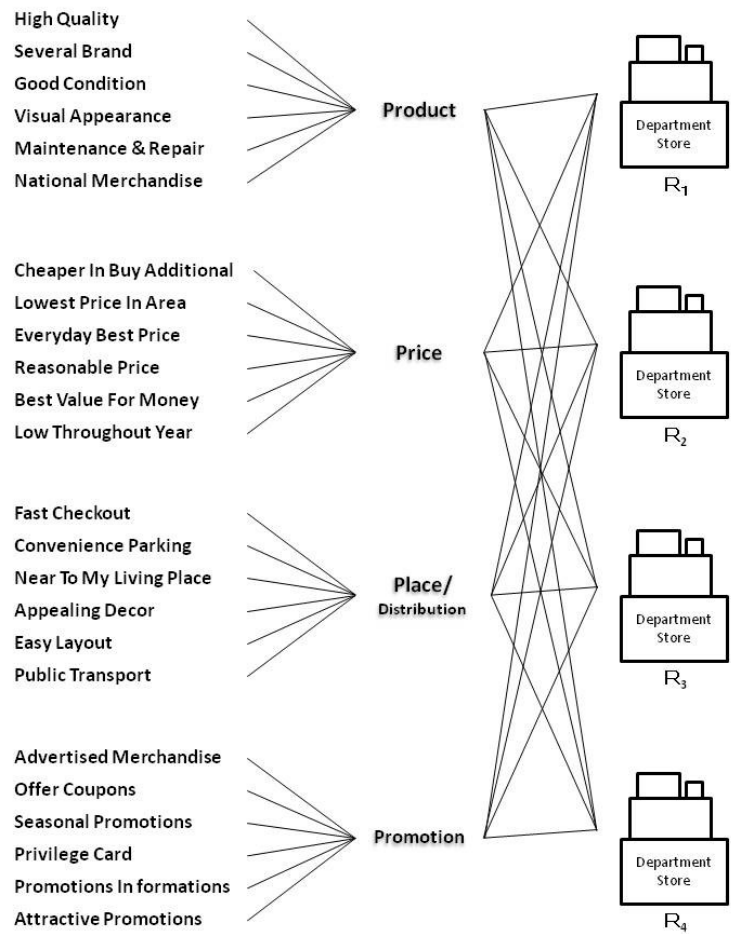

The illustration of Attribute - 4P's - Retail Stores Mapping in Figure 1 was built to sprout a better understanding on our study framework. Figure 1 elucidates the main idea of how we determine the targeted attribute of the 4Ps and generate it in the questionnaire to meet out objectives. The relationship between the marketing mix, 4ps with the criteria lies in each P element were clearly linking to the four selected retail stores.

When all are agreed on the category of criteria, to examine each alternative concordance to the attribute, we presuming that the options are known, it remains to complete the decision matrix. The assessment is generally independent of the aggregation procedure; it was due to the fact that examination theoretical counts on the posterior aggregation operation are generally ignored by the designers. The location of the respective alternatives or transforming a pair wise comparison into a numerical (normalized) scale as, for example, in the so-called "Analytical Hierarchical Process"(AHP) (Saaty, 1980).

The utilities of a prearranged option, in the structure of multi-attribute utility, regarding each attribute, are jointly cardinal. They have therefore to be jointly evaluated (Pomerol & Barba-Romero, 1993). The support of a Multicriteria Decision Making methodology should be very useful in the case considering the difficulty either to validate the probabilistic independence or to aid the decision maker to jointly measure the options by solvability or by the mid-preference point method.

## 3.4 DATA COLLECTION

The data were collected by means of questionnaire. First appointment was conducted with the personal in-charge in each retail store to request cooperation and approval for data collection and survey respond via formal letters from the Department of Mathematical Sciences, Faculty of Science and technology, National University of Malaysia.

Field research was conducted in Tesco Saujana Impian Kajang, Carrefour Alamanda Putrajaya, Giant Bukit Tinggi and Mydin Mart Kajang. A simple random sample of 214 household's respondents was obtained from each of the four retail stores; sum up a total of 856 respondents data.

## 3.5 Data Analysis and Interpretation

The retail market place promotes continuous improvement to survive in a turbulent atmosphere. For that, benchmarking is the exploration for industry best practices that leads to superior performance (Camp, 1989). The benchmarking dimension of the retail stores conceives a set of indicators and for this reason assumes the configuration of a multi-criteria analysis. The literature on retail stores and marketing mix model has identified four major underlying criteria essential to take place in the market place. They are as follows:

$ATT_1$ : Product Attribute

$ATT_2$ : Price Attribute

$ATT_3$ : Promotions Attribute

$ATT_4$ : Place/Distribution Attribute

An organization will show better performance on the basis of some indicators and worse performance on the basis of some others: "there is no single performance management enterprise system which is best in class across all areas" (Sharif, 2002).

Computed by averaging the scores assigned to all the organizations on the basis of all the criteria, we could obtain the result of the "best in class" in the organization, with the maximum averaged value.

Consider four retail stores:

$R_1$ : Tesco

$R_2$ : Mydin

$R_3$ : Carrefour

$R_4$ : Giant

The contribution of the multi-criteria outranking methodology to the valuation of the impact of marketing mix on customer satisfaction on four retail stores in terms of benchmarking analysis is significant. The application of outranking approach enables the benchmarking of the impact of marketing mix without the necessity of an aggregate indicator obtained by averaging all scores assigned to the organizations on the basis of the different criteria.

**3.6 Benchmarking and Outranking-Satisfying Methodology**

Developed by Operational Research, the outranking methodology is a family unit of algorithms (Roy, 1985; Vincke, 1992; Roy and Bouyssou, 1993; Pomerol and Barba-Romero, 2000). Of these, ELECTRE I method will be introduced here. The input of the ELECTRE I method is represented by a multi-criteria matrix as in Table 1, surrounded by a line containing the weights that the decision making assigns to each criterion.

Table 1 Multicriteria matrix (ELECTRE I)

|  | $ATT_1$ (Product) | $ATT_2$ (Price) | $ATT_3$ (Promotion) | $ATT_4$ (Place/Distribution) |
|---|---|---|---|---|
| $R_1$ (Tesco) | 4.42 | 3.94 | 3.97 | 3.90 |
| $R_2$ (Mydin) | 3.91 | 3.73 | 3.42 | 2.95 |
| $R_3$ (Carrefour) | 4.10 | 3.60 | 3.71 | 3.70 |
| $R_4$ (Giant) | 3.90 | 4.02 | 3.76 | 3.92 |
| Weight | 1/4 | 1/4 | 1/4 | 1/4 |

From Table 1, the retail stores' positioning is generated and shown in the table below:

Table 2 Retail stores Positioning Table

|  | $ATT_1$ | $ATT_2$ | $ATT_3$ | $ATT_4$ |
|---|---|---|---|---|
|  | (Product) | (Price) | (Promotion) | (Place/Distribution) |
| 1st | Tesco | Giant | Tesco | Giant |
| 2nd | Carrefour | Tesco | Giant | Tesco |
| 3rd | Mydin | Mydin | Carrefour | Carrefour |
| 4th | Giant | Carrefour | Mydin | Mydin |

Average ( $R_N$ ) = [ $ATT_1$ ( $R_N$ ) + $ATT_2$ ( $R_N$ ) + $ATT_3$ ( $R_N$ ) + $ATT_4$ ( $R_N$ )]/4

Now, let us consider $R_2$ and $R_3$. Taking into account the values in Table 1 it is evident that $R_3$ is better than $R_2$ for three criteria out of four (Marketing Model 4Ps). That is:

$ATT_1$ *( $R_3$ )* = 4.10 > $ATT_1$ *( $R_2$ )* = 3.91

$ATT_3$ *( $R_3$ )* = 3.71 > $ATT_3$ *( $R_2$ )* = 3.42

$ATT_4$ *( $R_3$ )* = 3.70 > $ATT_4$ *( $R_2$ )* = 2.95

Three criteria {1, 3, and 4} agree in considering $R_3$ better than $R_2$. Only one criterion {2} considers $R_2$ better than $R_3$. That is:

$ATT_2$ *( $R_2$ )* = 3.73 > $ATT_2$ *( $R_3$ )* = 3.60

Concordance-discordance principles are used to build outranking relations. Interpreting the same procedure for all the other pairs of retail companies will obtain the Table 3.

Table 3 Matrix of Concordance Subsystems ( $J^c$ )

|  | $R_1$ | $R_2$ | $R_3$ | $R_4$ |
|---|---|---|---|---|
| $R_1$ |  | {1,2,3,4} | {1,2,3,4} | {1,3} |
| $R_2$ | Ø |  | {2} | {1} |
| $R_3$ | Ø | {1,3,4} |  | {1} |
| $R_4$ | {2,4} | {2,3,4} | {2,3,4} |  |

The generic element $J^c$ ( $R_i$ , $R_j$ ) of the matrix of Table 3 is given by:

$$J^c ( R_i, R_j ) = \{j \in J \mid ATT_i ( R_i ) \geq ATT_j ( R_j )\}; \text{ where: } \mathbf{J} = \{1, 2, 3, 4\}$$

Taking into account the weights assigned to the various criteria, a concordance index can be calculated for each pair of company ( $R_i, R_j$ ):

$$C ( R_i ; R_j ) = \sum_{j \in J_c} K_j ;$$

Where: $K_j$ is the weight assigned to the $j_{th}$ criterion.

For example, for the pair ( $R_3, R_2$ ) we have:

$$C\ (\ R_3\ ,R_2\ ) = K_1\ +\ K_3\ +\ K_4\ = 1/4 + 1/4 + 1/4 = 0.75 \text{ (75 percent)}$$

We therefore have a majority of criteria of 75 percent in favor of $R_3$ with respect to $R_2$. Iterating the same procedure for other pairs or organizations, we obtain the concordance matrix of Table 4.

Table 4 Concordance Matrix

|  | $R_1$ | $R_2$ | $R_3$ | $R_4$ |
|---|---|---|---|---|
| $R_1$ |  | 1 | 1 | 0.50 |
| $R_2$ | 0 |  | 0.25 | 0.25 |
| $R_3$ | 0 | 0.75 |  | 0.25 |
| $R_4$ | 0.50 | 0.75 | 0.75 |  |

The concordance indicator in Table 4 varies between 0 and 1. It is equal to 1 only if there is unanimity or a majority of criteria that are 100 percent in favor of $R_i$ with respect to $R_j$. In order to decide on the superiority of one retail company with respect to another, the decision maker should set a concordance threshold $\boldsymbol{C}^*$. Generally, it is chosen to be a majority greater than or equal to 75 percent (simple majority tightened), that is: $\boldsymbol{C}^* \geq 0.75$ (75 percent).

Taking into account the database of Table 4 and the concordance threshold $\boldsymbol{C}^*$ we have the following concordance test:

$$T_c\ (\ R_i\ ,\ R_j\ ) = \begin{cases} 1 \text{ if } C\ (\ R_i\ ;\ R_j\ ) \geq C^* \\ \\ 0 \text{ if otherwise} \end{cases}$$

The results of concordance test are shown in Table 5.

Table 5 Outcomes of Concordance Test

|       | $R_1$ | $R_2$ | $R_3$ | $R_4$ |
| ----- | ----- | ----- | ----- | ----- |
| $R_1$ |       | 1     | 1     | 0     |
| $R_2$ | 0     |       | 0     | 0     |
| $R_3$ | 0     | 1     |       | 0     |
| $R_4$ | 0     | 1     | 1     |       |

The ELECTRE I methodology considers another step: the construction of discordance test in order to take into account of an excessive "distance" (dissimilarity) between the scores $ATT_j$ ( $R_j$ ) and $ATT_i$ ( $R_i$ ). The discordance test ( $T_d$ ) is fulfilled if the distance:

$$D ( R_j , R_i ) = \max [ ATT_j ( R_j ) - ATT_i ( R_i )];$$

does not exceed discordance threshold **D\***. In order to simplify the analysis we suppose that the test of discordance is fulfilled by all pairs ( $R_i , R_j$ ).

The outranking methods consists in examining the validity of the proposition "a outranks b". The concordance test "measures" the arguments in favor of saying so, but there may be arguments strongly against that assertion (discordant criteria). The "discordant voices" can be viewed as vetoes.

There is a veto against declaring that "a" outranks "b" if "b" is so much better than "a" on same criterion that it becomes disputable or even meaningless to pretend that "a" might be better overall than "b". The logic of the test of discordance is quite similar to that on which statistical tests are based. Here as well, conventional levels of significance, like the famous 5 percent rejection intervals, are widely used. The decision maker decides the discordance threshold, that is he decides whether a hypothesis must be rejected or not.

If the discordance test is not passed alternatives a and b are said incomparable. They are too different to be compared. Taking into account both the concordance and the discordance test we construct a binary outranking relation S. Given two generic retail companies ($R_i$, $R_j$) we say that $R_i$ outranks $R_j$ if and only if the concordance test ($T_c$) and the discordance test ($T_d$) are fulfilled, that is:

$$R_i \ S \ R_j \ \text{if and only if } T_c \ \text{and} \ T_d \ \text{fulfilled.}$$

Because we suppose that the discordance test ($T_d$) is passed by all pairs ($R_i$, $R_j$) the outranking relation S coincides with the outcomes of concordance test of Table 5. That is:

$$R_i \ S \ R_j \ \text{if and only if } T_c \ \text{fulfilled.}$$

The relation S may be represented by the graph of Figure 2.

Figure 2 Graph of S from Table 5 (C* $\geq$ 75 percent)

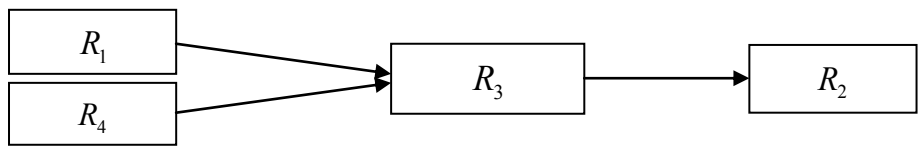

Now, $R_3$ is the "$2^{nd}$ worst in class" and $R_2$ is the "worst in class". But $R_1$ and $R_4$ are not comparable structures: neither $R_1$ outranks $R_4$ nor the opposite. This is another important difference arising from the refusal of the ordering based on the average benchmarking.

**3.7 Benchmarking On Customer Satisfaction**

Through benchmarking, we get better understanding of the customer because it is based on the reality of the market estimated in an objectivist way and a better economic planning of the purposes and the objectives to achieve in the company for they are centered on what takes place outside controlled and mastered. The management will get a better increase of the productivity, resolution of the real problems and understanding of the processes and what they produce.

FIGURE 3. Product Benchmarking towards customer satisfaction

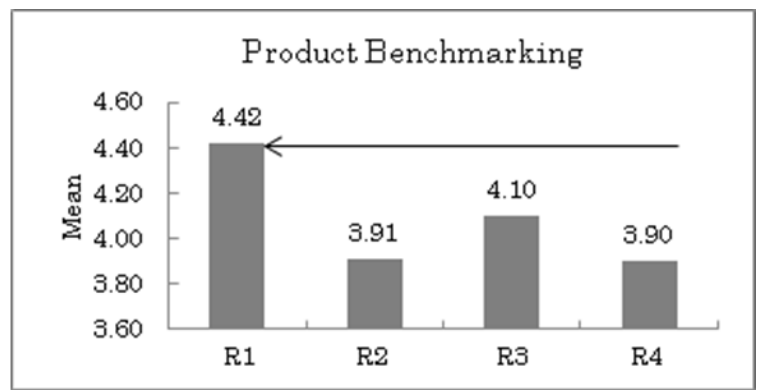

From Figure 3, $R_1$ ranks the highest on customer satisfaction towards product and it shall be the benchmark. $R_2$, $R_3$ and $R_4$ should benchmark $R_1$'s product strategy and improve to compete in the market.

FIGURE 4. Price Benchmarking towards customer satisfaction

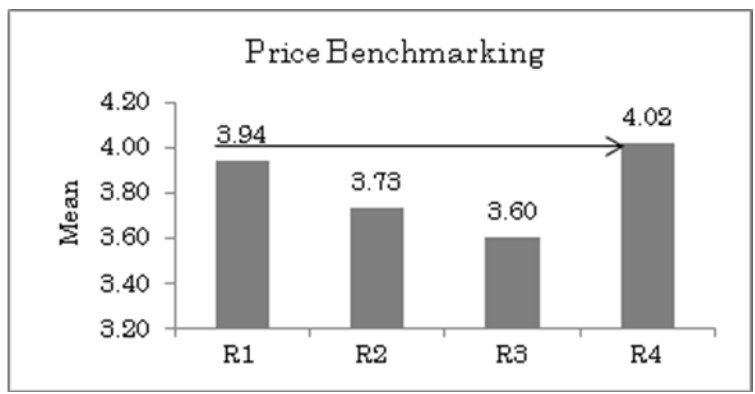

Figure 4 show that $R_4$ ranks the highest on customer satisfaction towards price. It proves that $R_4$'s "Everyday low price strategy" is a success. $R_1$ ranks the second with mean value of 3.94, in the competition mood with $R_4$. $R_2$ and $R_3$ should benchmark $R_4$'s pricing strategy.

FIGURE 5. Promotion Benchmarking towards customer satisfaction

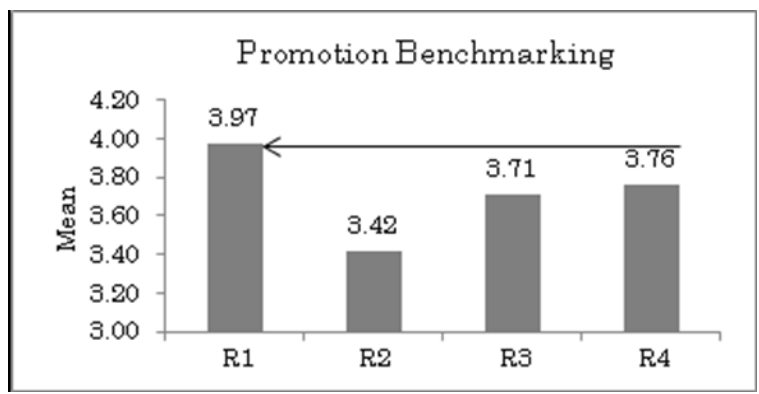

$R_1$ ranks the highest again in Figure 5 on customer satisfactions towards promotion, it is the benchmark. $R_1$ promotion strategy is well organized and effective; customers are aware of the latest promotion from the newspaper, flyers and promotion booklet. $R_4$ and $R_3$ are a little bit behind. $R_2$ ranks the last, it need to benchmark $R_1$'s and revise on its promotion strategy and improve to compete in the competitive market.

FIGURE 6. Place/ Distribution Benchmarking towards customer satisfaction

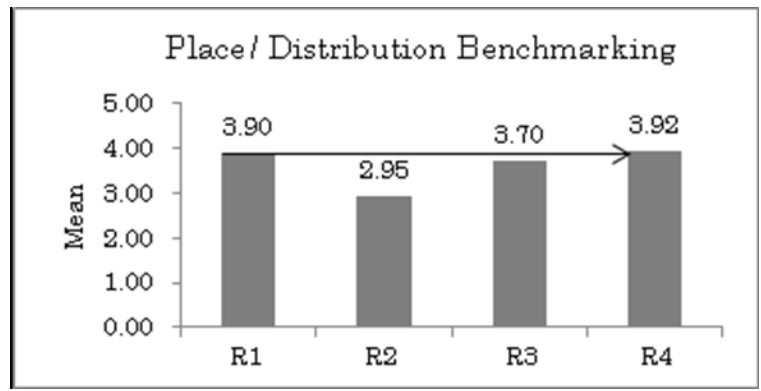

Figure 6 elucidates $R_1$ ranks the highest on customer satisfaction towards place and distribution and it shall be the benchmark for other retail stores. The other three retail stores having very close mean value. Meaning the customer satisfaction towards place and distribution in four retail stores are well perceived.

## 4. DISCUSSION ON SWOT ANALYSIS

It is not simply enough to identify SWOT of the ELECTRE I outranking method. Applying SWOT in this fashion can obtain leverage for a company (Ferrell, 1998).

### 4.1 Strength

The strength of MCDM is to aid decision-makers to be consistent with fixed 'general' objectives; to use representative data and transparent assessment procedures and to help the accomplishment of decisional processes, focusing on increasing its efficiency. The ELECTRE I method, in which the criteria of the set of decisional alternatives are compared by means of a binary relationship, often defined as outranking relationship, is more flexible than the ones based on a multi-objective approach.

**4.2 Weakness**

On a fuzzy angel of statement, often times different methods may yield different answers in terms of rankings when they are fed with exactly the same numerical data. It is a challenging and intriguing problem with decision-making methods which rank a set of alternatives practicing a set of number of competing criteria. Some kind of testing procedures need to be determined given that it is practically unworkable to know which one is the best alternative for a given decision problem.

**4.3 Opportunity**

In this paper, a new approach has been carried out for the use of the ELECTRE I model in marketing mix selection. This work shows that ELECTRE can be used successfully in deriving a consensus ranking in benchmarking to select the best in class.

**4.4 Threat**

In outranking approaches, the inaccuracy of the data can be modeled through the indifference and preference threshold, so-called pseudocriteria. Of course, threshold must be assessed for each criterion and for each problem separately.

**5. CONCLUSION**

As can be seen, the marketing manager should have rough outline of potential marketing activities that can be used to take advantage of capabilities and convert weaknesses and threats. However, at this stage, there will likely be many potential directions for the managers to pursue. The manager must prioritize all marketing activities and develop specific goals and objectives for the marketing plan (Boone, 1992).

It the effort of avoiding the shortcomings of the traditional methods based on the average aggregate monocriterion, outranking methods make it possible to deal with multicriteria benchmarking. They are a complete alternative to the traditional approach proviso applied to the measurement of learning capability. They can support the behavioral theory of organizational analysis initiated by H. Simon (Biggiero and Laise, 2003a, b). The behavioral theory is nonetheless perfectly comparable with them (Simon, 1997). The retail stores management uses the information so obtained to interpret the needs of individuals in the marketplace, and to create strategies, schemes and marketing plans.

The more the satisfying solutions will be when the lower the threshold assigned to the concordance test computing the lower the aspiration levels as a result.

## 6. DIRECTIONS FOR FURTHER RESEARCH

The relationships between customer satisfaction and behavioral outcomes are probably much more complex than initially assumed. This study has looked only at a limited part of the puzzle of how customer satisfaction translates into behavioral outcomes. In what way consumer characteristics moderate the relationship between satisfactions and repurchase behavior is likely to be contingent on the product or service category and the buying and usage process for that category. Other consumer characteristics not included in this study, such as a propensity for variety seeking behavior or a recreational shopping orientation, could potentially be important in many retail industries. Further research on how the effects of satisfaction on behavior is moderated by different consumer characteristics would advance customer satisfaction research as well as be of great managerial significance.

## ACKNOWLEDGEMENT

The authors are deeply indebted to National University of Malaysia for making this project a success. The authors express their gratitude to GCOE Meiji University for supporting on my ideas.    To fellow research assistants, Chen, Leong, Tan and Wong, much of this work and data collection was done in conjunction with them.